# A Lightweight Scrum Sprint Simulation to Help Learners Traverse the Empirical Process Control Threshold Concept

Eduardo Miranda
Department of Software and Societal Systems, Carnegie Mellon University
mirandae@andrew.cmu.edu

Torgeir Dingsøyr
Department of Computer Science, Norwegian University of Science and Technology *and* Department of IT Management, SimulaMet, Norway
torgeir.dingsoyr@ntnu.no

Pritam Chita
Department of Applied Informatics, Edinburgh Napier University
p.chita@napier.ac.uk

## Abstract

Empirical process control, a way of managing work based on the observation of the successes or misfortunes of earlier activities, is a key process in Scrum and other agile development frameworks. In this experience report, we present a lightweight, scalable, free and customizable sprint simulation activity designed to teach students how to empirically control a Scrum project by engaging in the presentation and interpretation of work status information, task selection and resource allocations in a single teaching session. We reflect on our experience using the simulation as an active learning complement to direct instruction in two master level courses at two different universities and in the training of teaching assistants at a third institution, and abductively establish its effectiveness by mapping student comments to the teaching practices in the threshold concepts framework.

## CCS Concepts

• **Software and its engineering → Agile software development**.

## 1 Introduction

Empirical process control underpins the adaptive and iterative nature of Scrum and other agile methods, relying on principles of transparency, inspection, and adaptation. These principles are operationalized in the following behaviors: updating project status indicators such as task boards, kanban boards, and burndown (or burnup) charts; interpreting them — e.g., analyzing trends, combining indicators, and making projections; deciding what tasks to tackle next, abandon, or preempt based on current progress, required effort left, resource availability, dependency fan-out and user value. In this paper, we describe a sprintsimulation activity and evaluate it through the lens of threshold concept theory [14], by mapping student comments to the teaching practices associated with threshold concepts traversal. This framework is appropriate because as argued in Section 2.2, we have found empirical process control to be a threshold concept in our teaching of agile methods. Our sprint simulation consists of two components: a simulation engine—comprising two lotteries or fortune wheels, a task board, burndown release charts, and an event log—and a simulation script, which is followed by the instructor to effectively guide the traversal process. The simulation embodies the execution of a Scrum sprint without coding and in virtual time, which allows it to be executed in a single teaching session. The entire simulation is free, lightweight, scalable, and customizable.[1]. Compared to project-based approaches [19], simulation approaches have the following advantages: time economy, reduced cognitive load, richer event coverage, collective learning, timely and situated feedback, and engagingness. Curriculum-wise, the simulation can be used by itself or as an active learning component of a direct instruction course. In any case, we assume that before the simulation starts, learners have declarative knowledge about Scrum ceremonies, roles and artefacts, and task allocation strategies and their consequences. The rest of the experience report is organized as follows: Section 2 (Background) introduces the use of simulation in the teaching of Scrum, a characterization of empirical process control as a threshold concept, and the methodologies to teach them. Section 3 describes the simulation engine, and Section 4 describes the simulation activity. In Section 5, the authors share their insights from using the simulation in practice and abductively establish its effectiveness. Section 6 provides a summary of the main contributions of the paper and the way forward.

## 2 Background

Because of its work management aspects, its mandatory reflection practices, its wide use in industry[2], and its deceptive simplicity[3], most introductory and advanced courses in software engineering adopt Scrum as a vehicle for teaching agile methods [3]. According to this review [3], project-based learning is the prevalent approach to teaching Scrum, while gamified learning, blended learning, and other experiential learning remain underutilized.

Project-based learning, particularly in capstone courses, provides students with opportunities to develop general industry-relevant skills [20]. Gamified learning instead provides domain-specific skills without project work time requirements. In a recent report, Di Nardo et al. [6] explore the advantages of gamification, concluding that gamification in education should be viewed as a valuable support tool rather than a total replacement for traditional teaching methods.

Preprint of article accepted for the software engineering education track at the ACM International Conference on the Foundations of Software Engineering 2026. The final version can be found here: https://doi.org/10.1145/3803437.3805776.

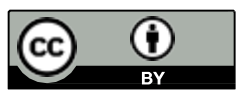

[1] A manual can be found at https://doi.org/10.5281/zenodo.19297363

[2] digital.ai, https://stateofagile.com/. Accessed January 2025.

[3] In the Scrum Guides from 2013 to 2017, Ken Schwaber and Jeff Sutherland, the creators of Scrum define it as a lightweight framework that is easy to understand but hard to master.

## 2.1 Scrum (Serious) Games

Christensen and Paasivaara [2] report the many different types of games that have been used to teach Scrum: cards, boards, balls, Lego bricks, and computer games. They also cite the creators of Scrum saying that "Scrum is simple to understand but difficult to master" and that it can only be learned "by doing".

In particular, lego4scrum [10] is a simulation with Lego bricks which originated in industry and is now used in many university courses. In lego4scrum, teams use Scrum practices to build Lego buildings for a city. A trainer plays the role of a product owner, who shares a vision and a product backlog. Teams of 4 to 6 people participate in sprint planning, work in "construction" during sprints, and conduct reviews and retrospectives.

The exercise ends with a retrospective on what happened during the three iterations the simulation lasts.

A study of an earlier version[4] of lego4scrum at Aalto University found that students learned about requirement management, customer collaboration, effective teamwork, and Scrum roles [16]. Students with varying backgrounds, including experienced developers, reported learning from the game.

The report [16] states that the game format is "simple" and Legos are "cheap tools", and while the students found some general industry-relevant benefits, the authors did not find that confidence increased in a project which followed the simulation. They stated that "the limited duration and scope prevent students from fully grasping all aspects of Scrum" and that students "missed the bigger picture of how the different Scrum practices complement each other" and "most importantly, students dismiss crucial Scrum practices as irrelevant or too cumbersome".

Note that many of these games aim to teach Scrum and life skills such as communication, teamwork, problem solving, time management, and self-directed learning. In our simulation, we focus specifically on empirical process control.

## 2.2 Threshold Concepts and Empirical Process Control

According to Meyer and Land [14], threshold concepts are core ideas within a discipline that are critical for mastering the field and hard to grasp. Without comprehending them, learners may replicate the language, behaviors, or actions associated with a concept without genuinely understanding its underlying principles or significance. From a learning perspective, these concepts are:

- *Troublesome*: Foreign, counterintuitive, complex, or contradictory.
- *Transformative*: Enable learners to think like experts.
- *Irreversible*: Once understood,
- *Integrative*: Reveal hidden interrelatedness.
- *Bounded*: Exist within disciplinary limits.

The literature on empirical process control, such as from the Scrum Alliance[5] describes it as a way of managing work through observation and experimentation based on the three pillars of transparency (visibility), inspection and adaptation. It is suitable for processes which are not well understood or where it is impractical to adopt a prescriptive, step by step approach. As stated in the introduction we operationalize this construct in terms of teachable behaviors: updating project status indicators; interpreting them; deciding what to do next and allocating resources accordingly. In Scrum, empirical process control is exercised through the following ceremonies: Sprint Planning, Daily Scrum meetings, Sprint Reviews, and Retrospectives. In our teaching practice we view empirical process control as a threshold concept because we frequently observe our learners struggling (troublesome knowledge) with things like: The need to break down user stories into pebble-sized tasks, hold daily meetings, track work progress, update burndown charts with planned hours instead of hours worked, pulling rather than pushing tasks, understanding the constraints of specialized team roles and internalizing the relation between work in progress and lead time. We believe these struggles might be due to the students' lack of experience working in group settings of more than three people or coming from cargo-cult [6] implementations of agile methods, and suddenly, in the middle of the simulation, having the epiphany (transformative experience) of appreciating how having pebble sized tasks are essential for burndown charts to be a faithful depiction of the work progress, the constraints that specialized roles put on task allocation strategies or how the practices fit together (integrative experience), e.g., the relation between daily meetings, pebble sized tasks and social loafing. We have also observed the persistence of the learnings (irreversibility) in follow-on courses, capstone projects and master theses, and in the case of the teaching assistants, as will be presented later, in how they explain the concept to their students. The notion of boundedness is a little bit fuzzier but does not seem as critical as the others [4]. Flanagan and Smith [8] illustrate it by the use of specialist terminology that acquires a meaning in one subject that clashes with everyday usage or its meaning in another discipline. For example, there is a frequently cited anecdote[7] by K. Schwaber, one of the inventors of Scrum, about the meaning of "empirical process control" in agile methods that conflicts with the meaning given in the context of continuous process industries where it originates.

## 2.3 Teaching Threshold Concepts

As indicated, threshold concepts cannot be learned by simply being told; they must be apprehended by the learner.

The time between the moment a concept is introduced—e.g., during a lecture—and when the learner internalizes it is called a "liminal space," "liminal period," or "liminal state." This is a time characterized by partial understanding and psychological anxiety induced by the conflict between old and new knowledge [12, 17]. Note that traversing the threshold is the desired outcome, but the literature also acknowledges the possibility of learners abandoning or withdrawing from the learning experience [1].

[4]Lego City Simulation is the common name used in many articles to refer to version 2, circa 2011 of the now called Lego4Scrum, version 3.

[5]See https://resources.scrumalliance.org/Article/empirical-process-controlscrum Accessed October 2024

[6]Replicating superficial aspects of a system, practice, or behavior without understanding its underlying principles or mechanisms.

[7]According to the anecdote, Ken Schwaber visited Babatunde Ogunnaike at DuPont Experimental Station in 1995 to discuss methodologies that were not working well in software development. What was learned from this meeting was the difference between defined (plan-driven) and empirical process control models from which its usage in agile development follows. This is different from the usage in Ogunnaike's book where he refers to the identification and construction of a process model from experimental data which is latter used to control a process.

Because learners need varying amounts of time and scaffolding to traverse the liminal state, a linear curriculum is often ineffective for teaching threshold concepts to a group. To address this, threshold concept theory proposes four tactics: recursiveness and excursiveness [11, 13], active engagement [1, 13], cooperative learning [9, 13], and reflective practice [5, 9].

- *Recursiveness and excursiveness*: A recursive approach revisits concepts from multiple perspectives, uses varied examples, and encourages learners to discover discrepancies between their current understanding and new evidence. Excursiveness allows deviation from planned paths to explore teachable moments and connect theory to practice.
- *Active engagement*: Learners must actively construct and restructure new knowledge based on prior learning. Instructors should prompt students to explain concepts in their own words, apply them in new contexts, and relate them to personal experiences. This can be achieved through real data, hands-on activities, and discussions.
- *Collaborative learning*: Students work in small groups, valuing each member's contributions. Listening, disagreeing, and expressing ideas are emphasized over simply finding the "right answer." This fosters engagement and creates a psychologically safe environment for intellectual risk-taking.
- *Reflective practice*: This involves examining thoughts, actions, and experiences to learn and improve. In teaching, it means connecting experiences to concepts, questioning what has been learned, identifying gaps, and forming hypotheses. Peer reflection introduces diverse perspectives and benefits from hindsight.

These four tactics form the framework we will later use to evaluate the simulation's fitness for use.

Differences in threshold crossing times also pose challenges for assessment. Therefore, low-stakes formative assessments are preferred over summative ones during the transition period.

## 3 Our Sprint Simulation

We describe our simulation as free, lightweight, scalable and customizable. Free because it is available under a Creative Commons license.[1] Lightweight as it uses standard classroom equipment, e.g. a projector, easel sized paper and sticky notes, and that the overhead introduced, in terms of instructor and student training and setup can be completed in less than an hour. Scalable because the time required to run it increases linearly with the number of members in a group, and not with the number of groups, and customizable because the simulation has few built in concepts which translates into total freedom to choose the events to be included, the rate of progress of the simulated tasks and the feedback given by the instructor.

During the simulation, learners:

- Update a task board, sprint, and release burndown charts with current project status.
- Make decisions about task selection and allocation.
- Compute effectiveness and efficiency metrics.
- Log events and decisions for later reflection.

The simulation runs in about two hours, with an additional hour recommended for setup and reflection. It consists of a simulation engine and a simulation script.

### 3.1 The Simulation Engine

The simulation engine is the mechanism used to drive the simulation. It consists of two lotteries or fortune wheels called the Events Wheel and the Progress Wheel (Figure 1), a task board, user and technical stories, a release burndown chart, a sprint burndown chart and a log to record significative conditions and decisions made, see Figure2.

The entries in the Events Wheel correspond to unplanned situations that arise in the life of a project such as "A critical defect is reported", "A new 10-points user story is requested" or "A team member must be absent for 3 days". As events are drawn, they must be addressed by each of the teams during their daily meetings. Instructors could easily delete existing events or introduce new ones to serve their teaching goals by modifying entries in the associated wheel forms.

The entries in the Progress Wheel denote individual daily progress hours. Each draw of the wheel will dictate the amount of planned work a team member was able to complete on a given day. For example, if a team assumes 6 hours of effective individual work each day, greater than 6 hours imply the team member achieved more than was planned, i.e. the work was easier than anticipated, entries lower than 6 hours per day denote work that was harder than expected, i.e. lower rate of progress. Our current wheel is calibrated to an expected 5.4 hours of progress with a standard deviation of 2.9 hours, which assures most teams won't be able to complete all planned work and will have to reprioritize some of it. As in the case of the Events Wheel, the mean value and the dispersion of the progress distribution can be changed by changing the value and/or the frequency of the entries in the wheel. Anecdotally, we have observed that extreme entries, e.g. 0 and 12 hours of progress and letting the wheel spin for a moment with a clattering sound helps build anticipation keeping participants engaged.

The task board, the release and a sprint burndown chart canvas, user and technical stories and tasks retain their customary Scrum definition and functionality. The log is just a diary for recording consequential decisions, see Figure 2, for consultation during the reflection session.

Learners might be asked to prepare sticky notes with user stories and tasks prior to the simulation, e.g., during a Sprint planning exercise, or the stickers might be generated and pre-printed by the instructor on self-adhesive labels. We use the first approach in our regular courses where the simulation is part of a longer running assignment and the second in workshops or other consulting interventions.

The stories have already been prioritized according to MoSCoW rules (Must have, Should have, Could have) and estimated in story points. Some user and technical stories have dependencies on other user stories. Tasks are associated with the user or technical stories they help realize with their work effort estimated in person hours. User stories and tasks are assumed to be carried out in a logical fashion, respecting their explicit dependencies, but as this is not wired into the simulation mechanics, the order could be relaxed if it serves a teaching purpose. For example, in one of the simulations, a team asked if it could bypass a user story creating an unplanned mock of it to save some time and deliver higher value to the user at the end of the sprint. Some tasks can only be performed by a particular team member, e.g. a specialist, while the rest of the tasks can be carried out by anybody. Specialist tasks are depicted in a different color or labeled as such.

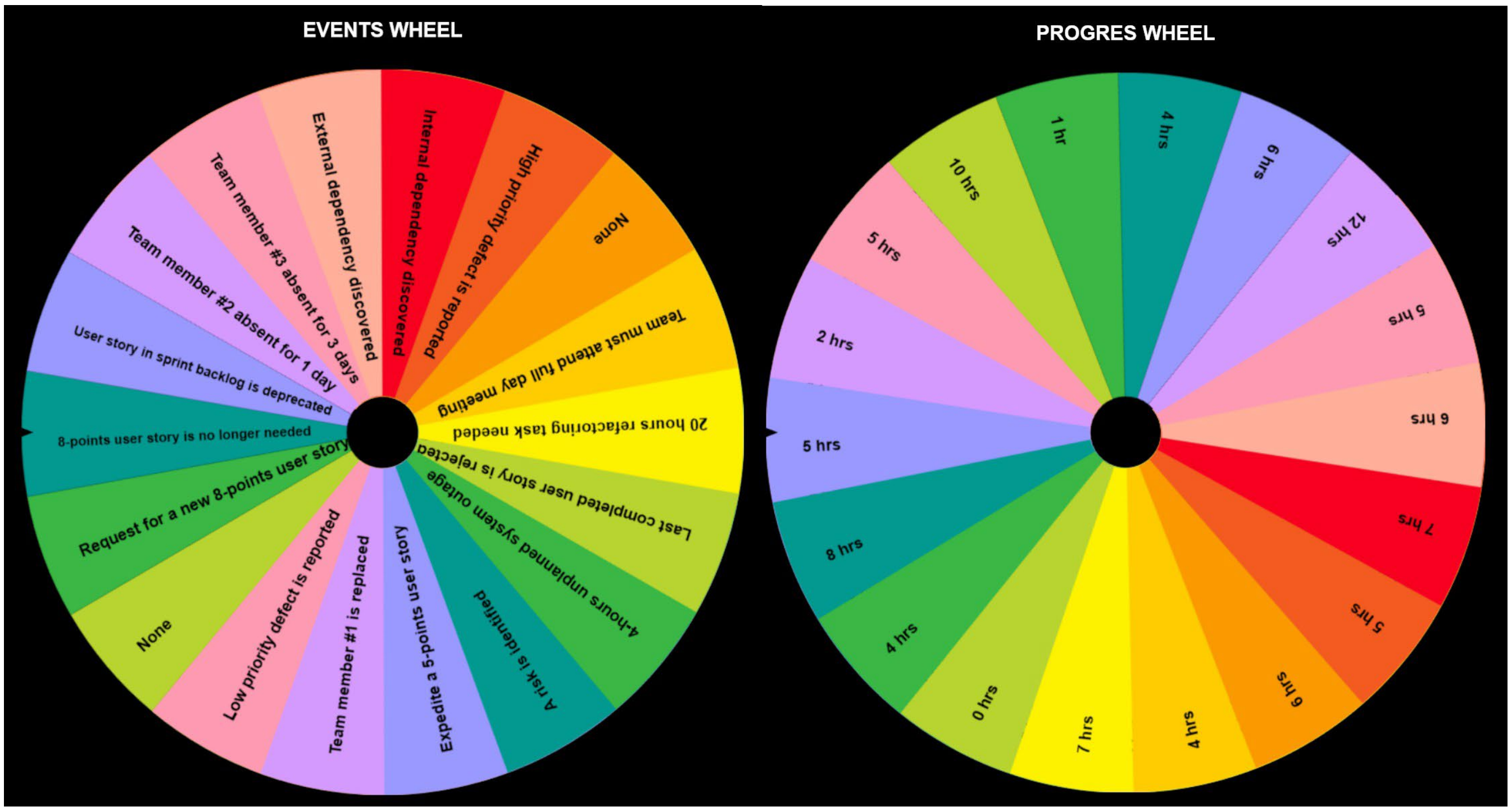

**Figure 1: The Events and Progress wheels.**

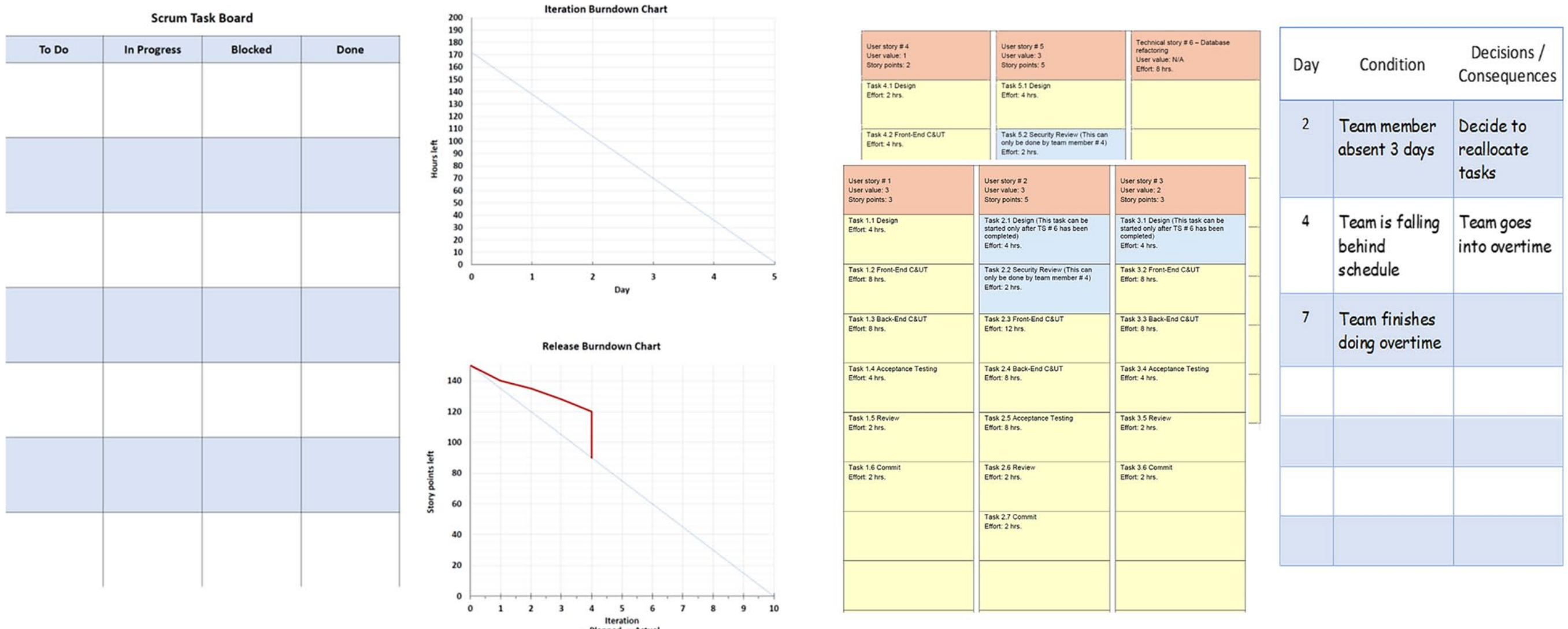

| Day | Condition | Decisions / Consequences |
|---|---|---|
| 2 | Team member absent 3 days | Decide to reallocate tasks |
| 4 | Team is falling behind schedule | Team goes into overtime |
| 7 | Team finishes doing overtime | |
| | | |
| | | |
| | | |
| | | |

**Figure 2: Task board, burndown charts, user and technical stories and tasks, decision log.**

In what follows, we describe the setup, the simulation workflow, and show how we have used the simulation to enable reflection.

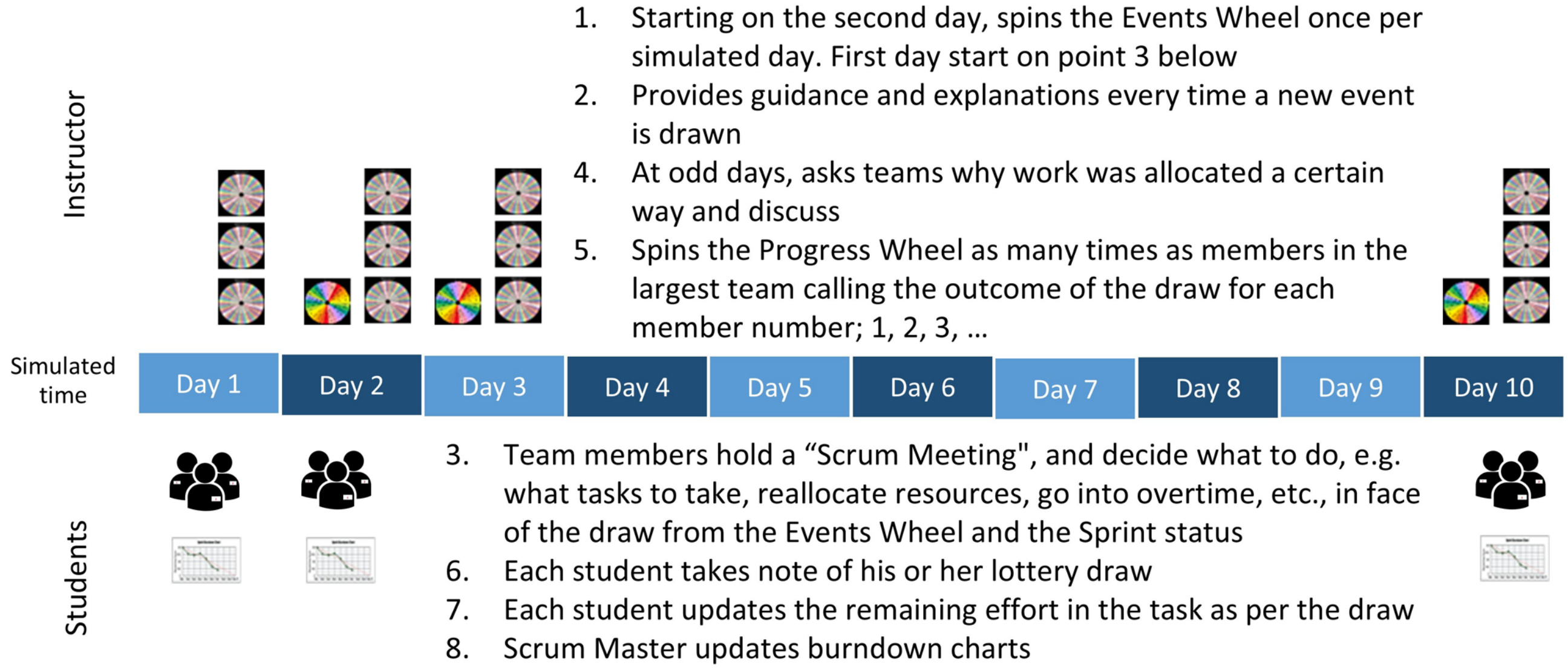

**Figure 3: Simulation workflow with tasks for the instructor and for the students.**

## 3.2 Simulation Script and Set-Up

Learners are assumed to have an understanding of Scrum and empirical process control prior to the start of the simulation; in our case this is done through lectures and other activities. For the simulation we recommend dividing a class into teams of four to six, with a preference for the first, and simulating no less than ten days to expose students to a broad range of events. These ten days can correspond to a two-week sprint or to two one-week sprints to demonstrate the transfer of incomplete or not accepted work from one sprint to the next. Each team will set-up a "station" on a section of the wall, consisting of the task board, the release and the sprint burndown chart canvas and the log, see Figure2. The task board is populated with user stories and tasks, and here comes the first challenge, drawing the ideal lines for the release and sprint burndown charts. Should the first include technical stories, should the second include the time spent in daily meetings, backlog refinement? What are the values of the ordinate at the origin? Since there are no authoritative answers to these questions, rather than arguing who is right and wrong, we prefer to ask what the consequences of the different choices are. This question invites deeper thinking and allows for the presentation and critique of multiple points of view.

Each team selects then a Scrum Master and assigns a consecutive identification number from one up to the number of people in the team to each member; it is advisable to write this number on a self adhesive name tag to avoid confusion during the simulation. As will be explained later, this number will associate each team member with their respective Progress Wheel draws.

An interesting and easy variation of the simulation is to create two types of teams, one composed of generalists who can perform a variety of tasks and another of specialists with specific roles, such as front-end developers, back-end developers, and testers and discuss the possible resource contention induced by the specialist organization. Before embarking in the full simulation, we like to run one or two practice draws to clarify the mechanics of the exercise.

*3.2.1 Simulation Workflow.* The simulation workflow for a 10-day sprint is depicted in Figure 3. The simulation starts with each team holding a Scrum meeting (3) to decide which tasks to tackle, dropping ongoing work, etc., who should tackle them, and whether to use overtime or not. Then the instructor spins the Progress Wheel, once for each team member (5), who use the amount of progress drawn, modified by overtime if applicable, to update the remaining effort on the tasks they are working on (6). Note the same draw applies to all team members with the same identification across all teams. This is the mechanism that allows the simulation to scale. So, other than by the larger number of queries and clarifications a bigger class will command, the wall clock time required to process each simulated day remains constant. The Scrum master updates the burndown chart (7). From the second day on, the instructor spins the Events Wheel (1) to draw an event which applies to all teams and must be addressed during the daily scrum meeting (3). Every other simulated day, or in response to previously unannounced events, the instructor asks the teams nudging questions. If he/she observes something that could be considered a teachable moment, it will pause the simulation to discuss it (2 and 4). To foster competition between teams, which promotes engagement, one can give a prize to the team which delivers the most user value at a minimum cost by the end of the iteration.

As mentioned above, teams might decide to use overtime to accelerate their progress. In such cases, the output from the progress wheel as it applies to each individual member will be multiplied by a factor to give the actual progress. Overtime hours are not as productive as normal hours, not only because as fatigue settles in, we do less, but also because we are likely to introduce more errors which need to be fixed later at a higher cost. Also, overtime is usually more expensive as it is paid at a higher rate. This combination of factors is weighed in a team efficiency metric which measures the amount of planned work achieved by every hour of labor put into the Sprint, see Figure5. Once more, this is not wired-in into the simulation, and it is the instructor´s choice whether to use it or not, and what weights to use to reflect the above dynamic.

*3.2.2 Simulation Reflection.* After the simulation part of the exercise is concluded we ask learners to reflect on it. We do this at the class and the team levels. The class-wide reflection involves all students present in the class and focuses on discussing the outcomes achieved by their respective teams—such as value delivered and costs incurred—along with the decisions made and the reasoning behind them (see Figure6). Additionally, we ask students about what have they learned and any "aha" moments they might have experienced, we also explore counterfactual scenarios by considering alternative actions and how these might have led to different outcomes. If teams used different team compositions, e.g., specialists versus generalists, we discuss its implications in terms of waiting times and bottlenecks. All this is part of a formative assessment for the exercise, which includes preparation, e.g. an open book multiple choice quiz about how the simulation works, the completeness of the material they bring to class, and their participation in the activity.

The team level reflection is a post-class activity students perform with the help of the simulation log to prepare an assignment report they need to submit for the summative assessment of the exercise.

## 4 Discussion and Lessons Learned

We now share our opinions on the benefits of using a simulation over a project-based learning approach in the context of our courses, how our sprint simulation is positioned vis-à-vis the lego4scrum simulation, and finally each of us relate their experiences in using the simulation.

### 4.1 Simulation vs. Project-based Learning

In our courses we chose simulation over project-based learning because of its:

- Time economy – The simulation is designed to be completed in hours instead of weeks.
- Process focus – The simulation eliminates the effort put into the construction of artefacts, e.g. code, not directly related to the course learning objectives.
- Discretionary event coverage – The simulation could include situations that will not arise naturally in a small, time bounded academic project, such as the need to refactor, requirements changes, the prolonged absence of a team member, changes in priority, etc., and therefore, never studied or discussed by the students.
- Collective learning – As will be explained later, because of its design, the simulation forces conflicting views within and between groups to surface, creating teachable moments.
- Timely and situated feedback – The context and decision making are visible to the instructor who could pause the work to provide clarification, clues and explanations and to ask for counterfactual scenarios.
- Engagement – The gamification of the simulation elements: lotteries, sound, teaming, manipulatives, and competition promotes excitement and receptivity.

This is not to say that simulation techniques should always be preferred to project based learning which clearly benefits the development of skills such as teamwork, time management, critical thinking and communication. Students often navigate open-ended problems that require integrating various knowledge areas. For example in their agile development through labs course Schroeder et al [18] state as learning objectives "creating a clean architecture, code comprehensiveness, and rigorous testing all of which are very important problems, although they are not algorithmically complex".

### 4.2 Simulation with lego4scrum

The big ideas for the lego4scrum simulation are [10]:

- Focus on the agile mindset.
- A product owner needs to have a big vision to be shared.
- Several (or many) teams will be working together to build a single integrated product matching the big vision.

The simulation includes a Pre-game phase comprehending the following goals: pitching the product vision, forming the teams, user story mapping, product backlog refinement, effort estimation and work prioritization; an In-Game phase which is about building and improving the product and covers the following activities sprint planning, daily meetings, sprint review, sprint retrospective and the actual building of the Lego city; last there is a Post-game debriefing session within and across teams and includes a game-recap, lessons learned and changing the world activities. The instructor is available as a product owner, with the simulation putting emphasis on the requirements work and deviations between expectations of the product owner and the team, the typical lack of collaboration between teams and ineffective communication and organization within the teams which typically lead to many tasks being started but few completed and approved by the product owner. The simulation further lends well to working on estimation of tasks with techniques such as planning poker.

Based on the description above, we can see that the lego4scrum and our simulation, each focus on different objectives, so one could use one or the other or both based on the time available and the learning objectives proposed.

### 4.3 Experience with the sprint simulation

In the following, we report on our individual experiences with the simulation in a master-level agile methods courses, and on the training of teaching assistants who work in an introductory course in software engineering at bachelor level:

*Master level training:* The first author started using the sprint simulation in 2016 to teach students in the intricacies of empirical process control [15], and while he always intuited the simulation was a key aspect of his course, it wasn't until he learned about threshold concepts, that he acquired the vocabulary to fully describe the what and the why of it. The sprint simulation is run near the end of a seven-weeks course on Agile Methods. The course's learning objectives are twofold: to cover the fundamental tenets of agile development, he calls this the course enduring principles, and to develop immediate competence in a concrete method, in this case Scrum. The course is taken by between 30 and 40 students and involves two weekly lectures, and a mandatory weekly class activity performed in groups of three to six students.

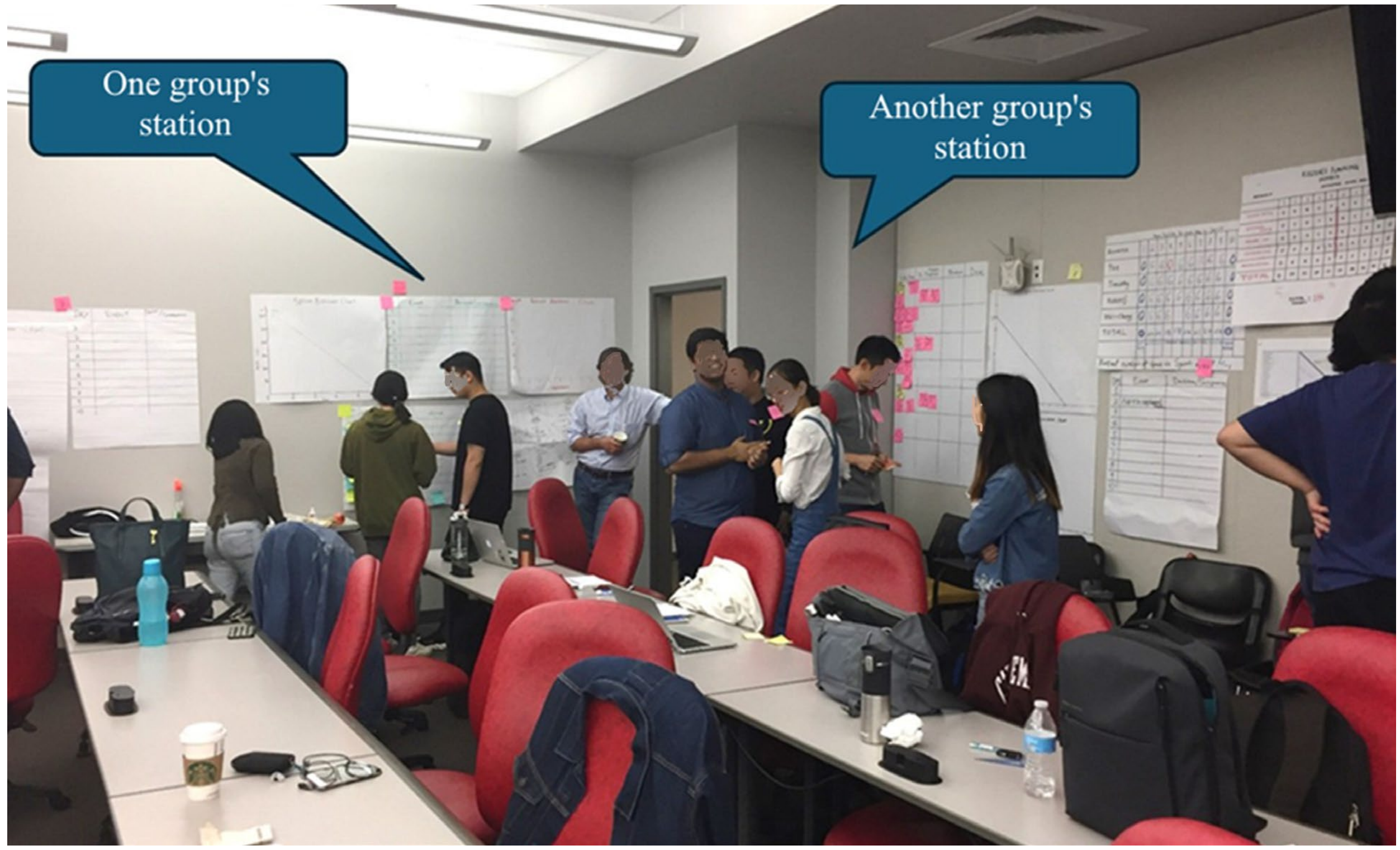

**Figure 4: Groups' stations.**

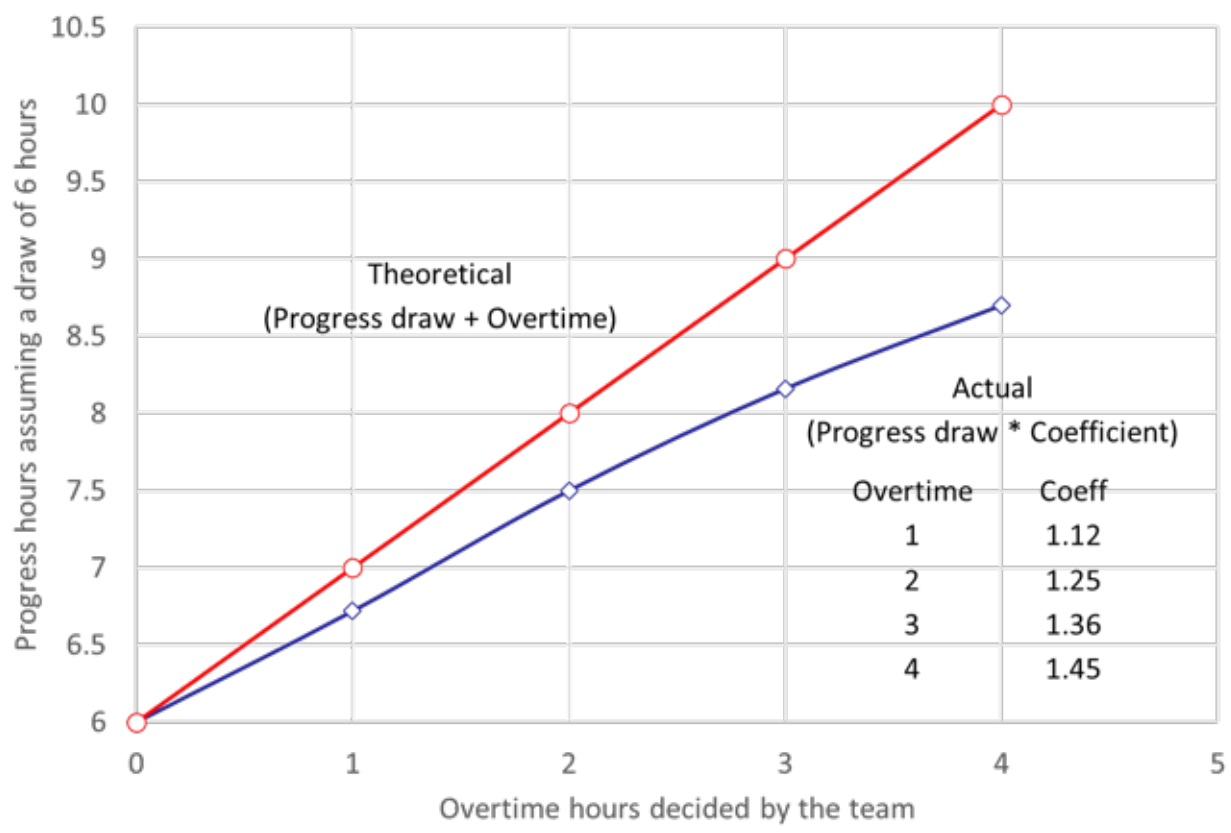

**Figure 5: Overtime hours are not as productive as regular hours.**

Very early in the use of the simulation, he noticed that a large number of students were having an "Aha moment" during or just shortly after participating in it. That moment, the time when all the concepts taught during the lectures came together and the students could genuinely make sense of them, was the manifestation of a learner having traversed the liminal space associated with the empirical process control concept. This "Aha moment" is illustrated in a quote from a student (from simulation by third author), "What I valued most was the ability to simulate unpredictability through the events and the progress wheels".

A retrospective analysis of the simulation through the lenses of threshold concept theory, see Table1, has given him confidence in the soundness of the teaching approach and the means to justify it to other colleagues.

The third author ran the simulation in early 2025 on a Masters level "Management of Software Projects" module centred on the

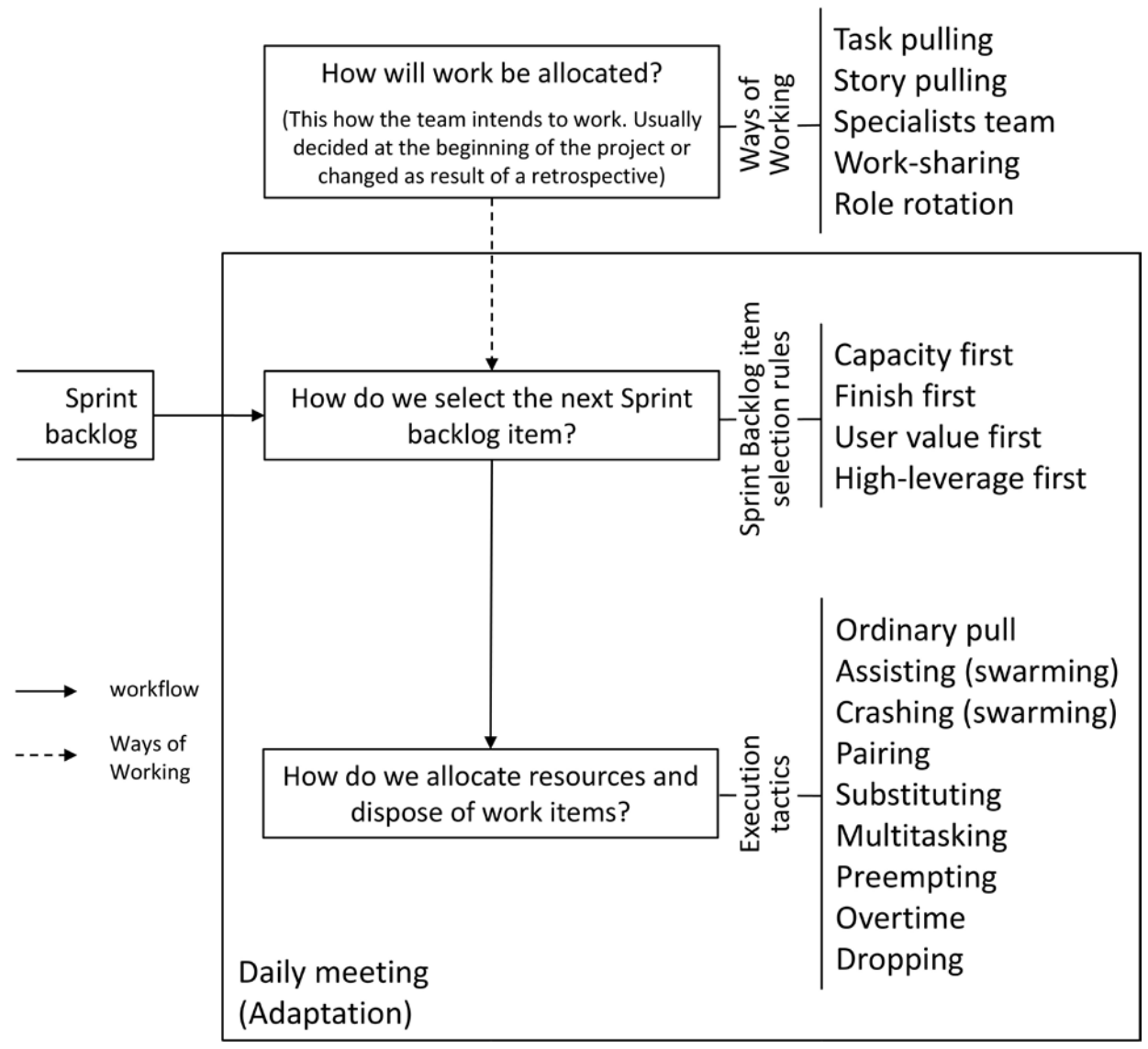

**Figure 6: Ways of working and strategies for task allocation.**

Scrum method. Previously a test version of the Scrum Card Game (https://agilelean.pro/scrum-card-game/) was used but students had difficulty in translating game playing elements to those of Scrum practice. The simulation was run in week 9 of a 15 week semester and was used as a means of coalescing multiple teaching points. With a cohort of 106 students, the simulation was run in four two hour tutorial groups with over 25 students divided into groups of 5. Student engagement during the simulation was very high with students indicating they were activity involved in discussions as well as manipulating the simulation artifacts such as task boards and charts. The real-time feedback afforded by the simulation was regarded as extremely valuable with most students reflecting on previous simulation decisions and making adjustments as the simulation progressed. This was evident from students comments. "What I enjoyed most about the sprint simulation exercise was how realistic and interactive it was. I learned to experience working in an Agile environment where collaboration and flexibility take center stage . I enjoyed how it forced collaboration seeing how tiny conversations and quick decisions could have a profound impact was an eye-opener." In addition, students had exposure to the real world uncertainties of project delivery and the need for constant re-evaluation. "What I valued most was the ability to simulate unpredictability through the events and progress wheels. This taught us the importance of adaptability and contingency planning as we had to reallocate tasks and manage new story points while completing planned story points"

*Training of teaching assistants:* The second author used the simulation to train 24 teaching assistants who would play roles as product owners and supervisors of a total of 72 teams of students who take a bachelor-level introductory course in software engineering [7].

The teaching assistants stated learning outcomes such as being "reminded that unforeseen events often happen, and how much it can affect a workday", "A lot can happen unexpectedly, difficult to plan when it is uncertain how much time you have", but also aspects such as "the importance of collaboration and being mentally prepared for change" and "distribution of work in a context with dependent user stories", and that a team "always has something to do to maximize efficiency". Half a year after running the simulation, some teaching assistants stated that this was central to their understanding of Scrum, they had not realized until then the importance of the decisions taken in the daily meetings on distribution of tasks.

## 4.4 Lessons learned

We summarize the main lessons learned from using the simulation as a training technique in various courses for several years:

- The simulation has a number of advantages to project-based learning in time economy, process focus, discretionary event coverage, collective learning, timely and situated feedback and in providing engagement.
- Our simulation specifically focuses on empirical process control which is a key concept in Scrum, unlike other games as lego4scrum which have different learning objectives.
- Student feedback is overall positive, and we illustrate how the training method aligns with recommended tactics to overcome threshold concepts: Recursiveness and excursiveness, active engagement, collaborative learning and reflection (See Table 3).
- Based on our experience, we recommend the simulation for training of students at any level given that they already have a basic understanding of Scrum.

# 5 Summary

We began our experience report by operationalizing empirical process control by means of teachable behaviors and describing our free, lightweight, customizable, and scalable simulation. We then introduced threshold concepts theory and its core teaching practices: recursiveness and excursiveness, active engagement, collaborative learning, and reflective practice and used this framework to hypothesize the effectiveness of our simulation by interpretively mapping students' comments to said practices. We compared our simulation to project-based learning and to the well-known Lego4Scrum simulation concluding ours is time economical, focused, allowing instructors to control the context and ensuring that the process aspects remain the primary learning focus. It is engaging and integrative, and its interactive nature mitigates free riding and fosters teamwork. Furthermore, it enables instructors to provide timely, situated feedback that promotes reflection-in action. We are following this experience report with a survey of our students across three different institutions to confirm our preliminary conclusions.

**Table 1: Recommended teaching tactics for threshold concepts and their implementation through simulation elements along with student comments evidencing their effectiveness.**

| Recommended tactic | Simulation elements implementing the tactic | Student comments regarding the experience |
|---|---|---|
| Recursiveness | • Daily meetings (repetition)<br>• Event wheel (new takes)<br>• Progress wheel (induces new takes via uneven progress in the tasks) | • Some topics are difficult for me to understand during the lecture, but the class activities helped me to get to know them a lot<br>• . . . the class activities help us reinforce what was taught in the class<br>• The simulation helped me understand the real-world aspect of this subject |
| Excursiveness | • Instructor prompts (Scripted and opportunistic interventions create teachable moments)<br>• Other group questions (opportunistic teachable moments)<br>• Own group questions (constructive conflict) | • . . . activities were a wonderful way to test your understanding of the theory from previous lectures<br>• . . . forced me to rethink what I thought I knew coming into the course |
| Active engagement | • The simulation itself<br>• Manipulation of physical artifacts (task board, charts, task and user stories)<br>• Gamification (lottery draws, lottery wheels sound) | • With the class activity, I can practice the concepts I have learned and deepen my understanding of them |
| Collaborative learning | • Small groups (4 to 6 members per team)<br>• Preparation session (requires group to decide initial values for burndown charts, and organize the task board)<br>• Post activity session (requires the group to agree on a write-up)<br>• "Daily" meetings (decisions on how to select tasks) | • As an MBA student, most of the material was completely new to me and I enjoyed learning about agile methods and practicing them in groups during class<br>• The activities in this class are great! They help me understand the agile methods better and build relations with others<br>• . . . taught me some of the methodologies that I had not experienced in my company but wanted to! |
| Reflection | • Instructor prompts (Reflection-in-action)<br>• Post activity session (Reflection-on-action)<br>• Performance metrics review (interpretations, consequences) | • I have worked in Scrum before so the course really worked towards reconciling what I thought I knew<br>• It is unique and I gain a huge amount of new information and correct some incorrect things I understood incorrectly about agile<br>• … forced me to rethink what I thought I knew coming into the course. |